\pdfoutput=1
\documentclass[final,5p,times,twocolumn]{elsarticle}

\usepackage{amssymb}
\journal{Physica C}

\begin{document}

\begin{frontmatter}



\title{BiOCuS: A new superconducting compound with oxypnictide - related structure}


\author[label1]{A. Ubaldini, E. Giannini, C. Senatore, D. van der Marel}

\address[label1]{D\'{e}partement de Physique de la Mati\`{e}re Condens\'{e}e, University of Geneva, 24 quai Ernest-Ansermet,
CH-1211 Geneva, Switzerland}

\begin{abstract}
The discovery of about 50 K superconductivity in the tetragonal Fe-based pnictides has stimulated the search for superconductivity in a wide class of materials with similar structure. Copper forms compounds isostructural to LaOFeAs. Single phase BiOCuS can be prepared by a solid state reaction at temperature lower than 500$^{\circ}$C from a mixture of Bi$_{2}$O$_{3}$, Bi$_{2}$S$_{3}$ and Cu$_{2}$S. The samples have been characterized by means of EDX analysis, X-ray diffraction, magnetic and electrical measurements. The cell parameters are $a$ = 3.8708 {\AA}, $c$ = 8.565 {\AA}. Charge carrier doping can be realized either by F substitutions for O, or by Cu off-stoichiometry. The latter doping route leads to the occurrence of superconductivity below T$_{c}$ = 5.8 K.

\end{abstract}


\end{frontmatter}


\section{Introduction}
\label{1}
The recent discovery of four families of iron-based superconductors: REOFeAs, AFe$_{2}$As$_{2}$, LiFeAs and Fe(Se,Ch) (where RE is a light rare earth element, A an alkaline earth or alkaline element, and Ch a chalcogenide: S or Te) has attracted a very large interest \cite{Kamihara08}. The $T_{c}$ of these materials can be as high as $\sim$ 56 K for REOFeAs \cite{Ren08}. Unfortunately, these materials contain highly toxic elements, particularly arsenic that is the most common cause of acute heavy metal poisoning in adults \cite{atsdr}. Besides, not all the physical aspects of these new superconductors are completely understood. The search for new materials with related structures and similar chemical properties, but without As, is very useful in order to investigate the physical properties of the iron-based pnictide superconductors.
Many compounds of metals with appropriate anions crystallize with the same structure as these new superconductors. For example, CaNiGe \cite{Hlukhyy08} and NaAlSi \cite{Kuroiwa08} are isostructural to LiFeAs. The second one is a superconductor ($T_{c}$ = 7 K). Silver forms compounds isostructural to LaOFeAs with S and Se \cite{Palazzi81}. This suggests that some other compounds isostructural to LaOFeAs may exhibit superconductivity if properly doped. The aim of the present work is to look for new tetragonal materials with copper, that has several stable valence states, and sulphur. Such quaternary compounds exist, either containing lanthanoid elements \cite{Hiramatsu08} or bismuth \cite{Hiramatsu08,Kusainova94}. The latter has revealed to be easier to prepare as a single phase, thanks to the higher reactivity of bismuth oxide and sulfide compared to those of lanthanum. Superconductivity is found in the compound BiOCu$_{1-y}$S.

\section{Experimental details}
\label{2}

The compound BiOCuS is synthesized by solid state reaction starting from stochiometric mixtures of Bi$_{2}$O$_{3}$, Bi$_{2}$S$_{3}$ and copper sulfide. Two stable copper sulphides exist: CuS and Cu$_{2}$S. Both can be used to prepare BiOCuS, but the use of the latter enhances the phase purity of the final sample. Therefore the chemical reaction used to prepare this oxysulphide is:\\

\begin{math}
    \frac{1}{3}Bi_{2}O_{3}+\frac{1}{6}Bi_{2}S_{3}+\frac{1}{2}Cu_{2}S \longrightarrow BiOCuS
\end{math}\\

\begin{figure}[h!]
\begin{center}
\includegraphics[width=0.9\columnwidth]{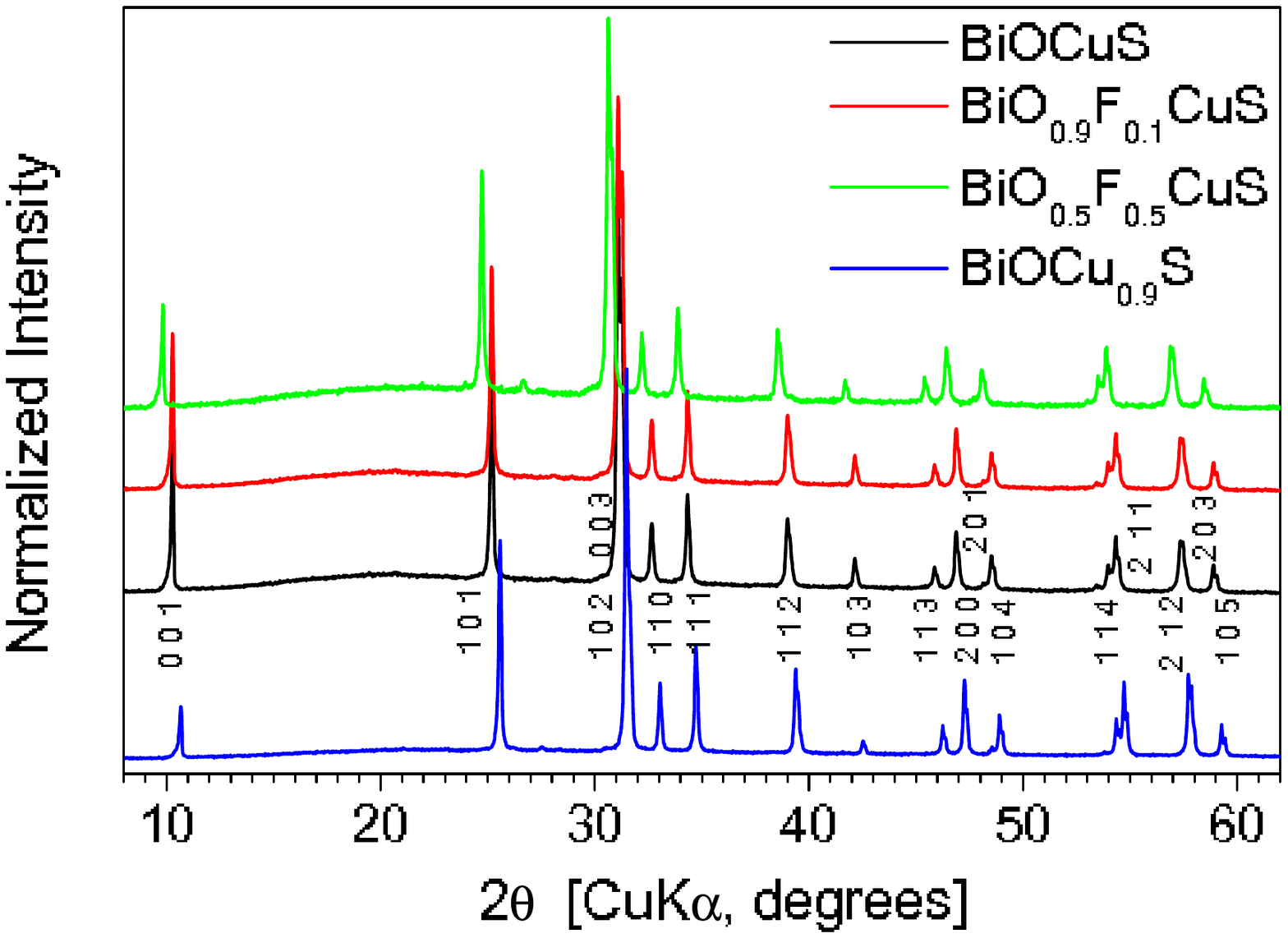}
\caption{XRD of pure and doped BiOCuS.\label{fig:figure1}}
\end{center}
\end{figure}

The precursors were carefully mixed, sealed in closed quartz ampoules under a low argon pressure and calcined several times between 400$^{\circ}$C and 800$^{\circ}$C for 50 hours each. LaOAgS, LaOFeAs, and LaOCuS would need higher processing temperatures \cite{Kamihara08,Palazzi81,Hiramatsu08}. Single phase material was obtained at 500$^{\circ}$C, after preliminary calcination at 400 and 450$^{\circ}$C with intermediate grinding (see Fig.\ref{fig:figure1}). BiOCuS starts loosing Cu and decomposes at T $\geq$ 650$^{\circ}$C. The cell parameters of the parent compound are a = 3.8726 {\AA} and c = 8.5878 {\AA}, in good agreement with previous reports \cite{Hiramatsu08,Kusainova94} .

In LaOFeAs superconductivity can be induced by either hole or electron doping, by partial substitution of La$^{3+}$ by Ce$^{4+}$ or O$^{2-}$ by F$^{-}$, respectively. A similar approach can be tried for BiOCuS. However, because of the high stability of CeO$_{2}$ at such low temperatures, the attempts to substitute Ce for Bi were not successful. As an alternative route for modifying the valence state of copper, we have controlled its content to be slightly deficient. Therefore, doped samples were prepared according to the reactions:\\

\begin{math}
\frac{1-x}{3}Bi_{2}O_{3}+\frac{x}{3}Bi+\frac{x}{3}BiF_{3}+\frac{1}{6}Bi_{2}S_{3}+\frac{1}{2}Cu_{2}S \rightarrow BiO_{1-x}F_{x}CuS
\end{math}\\

\begin{math}
\frac{1}{3}Bi_{2}O_{3}+\frac{1}{6}Bi_{2}S_{3}+\frac{1}{2}CuS+\big(\frac{1}{2}-y\big)Cu_{2}S \rightarrow BiOCu_{1-y}S
\end{math}

\section{Results}
\label{3}

In the case of BiO$_{1-x}$F$_{x}$CuS, nearly single phase samples were obtained up to x close to 0.5 by reacting at 490$^{\circ}$C during 100 h (see Fig.\ref{fig:figure1}). Large amounts of secondary phases, up to 20\% in volume, form when x $\geq$ 0.5. Fluorine losses accompanied by the formation of Bi$_{2}$O$_{2}$S and traces of Bi$_{2}$CuO$_{4}$, and rarely metallic Bi, occur when increasing the nominal F composition above x = 0.5. The cell volume follows Vegard's law as a function of the doping level. The $a$ axis slightly increases and the $c$ axis decreases when x varies from 0 to 0.5 ($a$ = 3.8785 {\AA} and $c$ = 8.5735 {\AA}, at x = 0.5).We have established using EDX, that part of the oxygen is substituted by fluorine.
However, the actual fluorine concentration in the BiO$_{1-x}$F$_{x}$CuS phase was observed to be lower than the nominal value (x=0.3 instead of x=0.5). It is expected that doping with fluorine in BiO$_{1-x}$F$_{x}$CuS modifies the charge of the Bi-O layer, thus inducing negative charges onto the Cu-S one. This can cause the formal reduction of the copper valence to (1-x). Interestingly, the magnetic susceptibility of these samples changes as a function of x, showing that the originally diamagnetic parent compound becomes paramagnetic at low values of x. The magnetic susceptibility follows a modified Curie law $\chi = \chi_{0}+C/T$, where $\chi_{0}$ is a temperature independent parameter. No evidence of either superconductivity or magnetic ordering was found down to 2 K.

\begin{figure}[h!]
\begin{center}
\includegraphics[width=0.9\columnwidth]{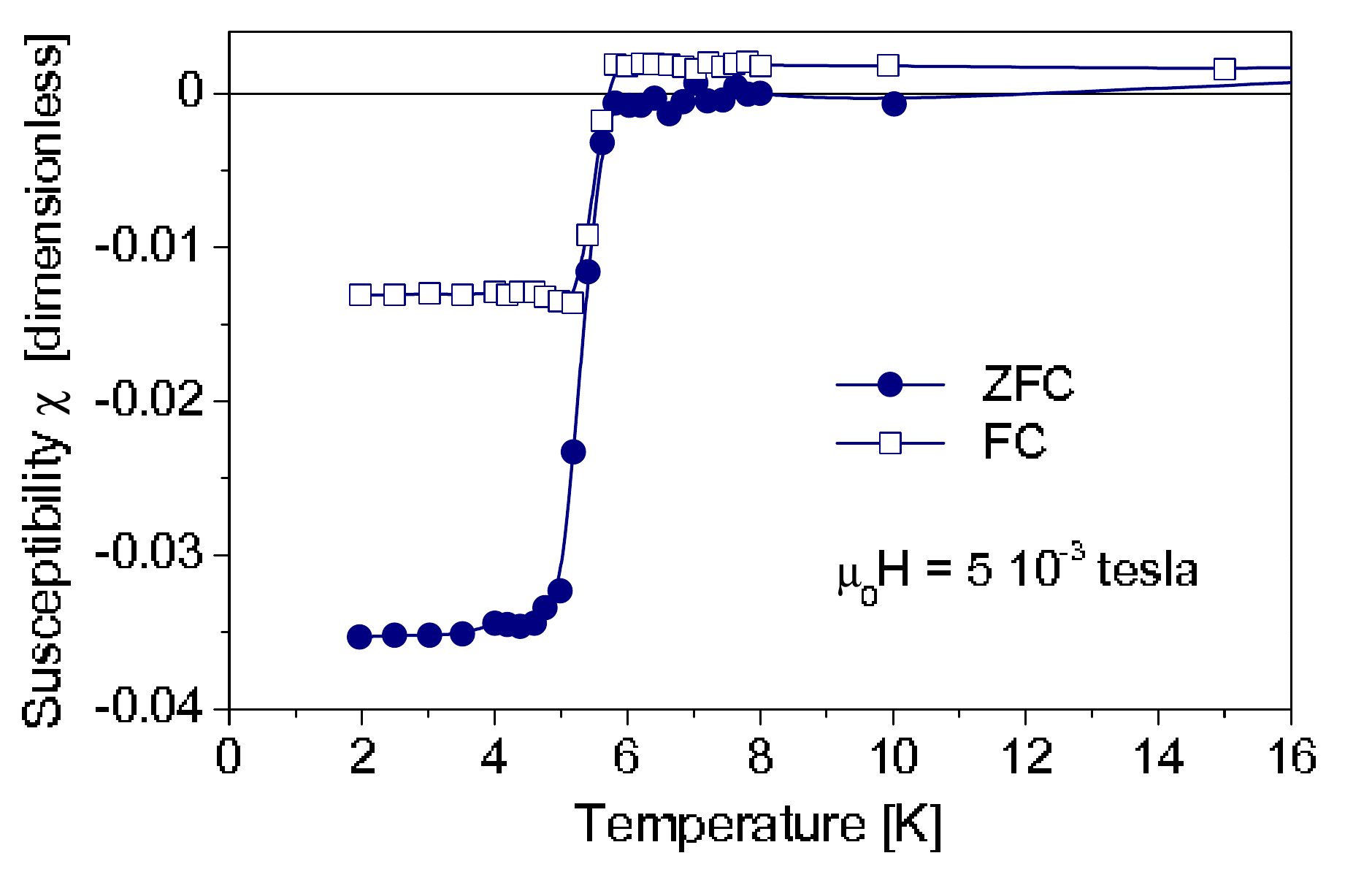}
\caption{Superconducting transition observed in the magnetic susceptibility. The low absolute value indicates that only a small fraction of the sample's volume is shielded by the supercurrents.\label{fig:figure2}}
\end{center}
\end{figure}

The second way for doping the BiOCuS is the control of the copper content. At relatively low level of off-stochiometry, up to y $\sim$ 0.15, single phase samples of BiOCu$_{1-y}$S can be prepared, at a reaction temperature of 520$^{\circ}$C. In this case, the average formal valence of copper is expected to slightly increase, in order to keep the electroneutrality of system. Because of the low sensitivity and limited reliability of the EDX analysis in measuring the Cu deficiency, the change of the lattice parameters is taken as the proof that the stoichiometry has changed ($a$ = 3.8686 {\AA} and $c$ = 8.5682 {\AA}, at y = 0.1. See Fig.\ref{fig:figure1}). Very interestingly, and even surprisingly, in the BiOCu$_{0.9}$S sample a superconducting transition is observed at 5.8 K, as shown in Fig.\ref{fig:figure2}. The low intensity of the magnetic shielding (4\%) leaves the question open, whether superconductivity occurs in the bulk or locally at the surface and/or in some small, granular and not connected volume.
The observed transition cannot be due to any secondary phase or non reacted precursors: among them, pure Bi can only become superconducting under pressure, and CuS is superconducting below 1.6 K \cite{DiBenedetto06}. The field cooled susceptibility is 1.6\%, providing the strongest evidence for superconductivity below 5.8 K. A superconducting hysteresis loop opens in m(H) measurements and is shown in Fig.3. We speculate  that superconductivity is limited to a composition range much narrower than the natural spread of y in this material.

\begin{figure}[h!]
\begin{center}
\includegraphics[width=0.9\columnwidth]{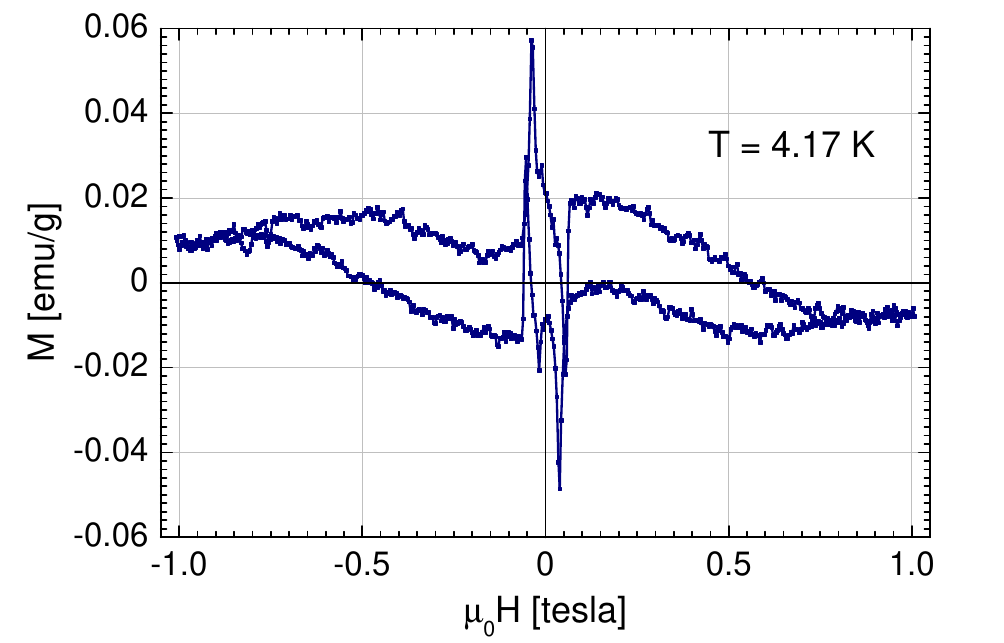}
\caption{Hysteresis loop of BiOCu$_{0.9}$S. Flux jumps appear at low field. \label{fig:figure3}}
\end{center}
\end{figure}


\bibstyle{elsart-num}
\bibliography{Ubaldini09_1}


\end{document}